\documentclass[aps,prd,preprintnumbers,nofootinbib]{revtex4}
\usepackage{bm}
\usepackage{graphicx}
\newcommand{\bea}{\begin{eqnarray}}
\newcommand{\eea}{\end{eqnarray}}
\begin{document}
\preprint{arXiv:1301.2561v2[nucl-th]}
\title{Anomalous 'mapping' between pionfull and pionless EFT's}
\author{Ji-Feng Yang$^{\dag,\ddagger}$}
\affiliation{$^\dag$Department of Physics, East China Normal
University, Shanghai 200241, China\\$^\ddagger$Kavli Institute for
Theoretical Physics China, Chinese Academy of Sciences, Beijing
100190, China}
\date{\today}
\begin{abstract}
The pion contributions to the coupling $C_0$ of pionless EFT are
studied via both non-relativistic and relativistic forms of chiral
effective field theory for nuclear forces. A definite item in the
$2N$-reducible component of the box diagram is shown to be dominant
over the $2N$-irreducible (potential) ones due to the pinching of
low-lying nucleon poles, and this anomalous mapping between pionless
and pionfull EFT's occurs right within the non-relativistic regime.
A natural strategy for renormalization of the pionfull theory
emerges as a byproduct through the interactive analysis of the box
diagram. Such mapping perspective may shed some light on the
efficient organization of the pionfull effective field theory for
nuclear forces.
\end{abstract}
\maketitle
\section{Introduction}Pions are the first particles known to mediate
strong interactions between nucleons. After quark picture of hadrons
is established, they are degraded as effective degrees of QCD. For
nuclear forces, however, the direct analytical computation using QCD
is in fact impossible, hence effective theories are extremely useful
tools at hand. Since Weinberg's seminal work in 1990\cite{Weinberg},
there have been great progresses in applying EFT methods to
nucleonic systems\cite{BBP,BvK,epelrev,EMRMP,EM_RP503,EM12,Epel13},
pretty laying down the field theoretical foundation for nuclear
physics. Intriguingly, there still remains an unsettled issue that
is concerned with the nonperturbative treatment of pion-exchange
potential\cite{NTvK,PVRA1,PVRA2,PVRA3,EpelMeis1,EpelMeis2}. Two
prevailing choices are adopted in literature concerning this issue:
(1) Nonperturbative treatment\cite{EM1,EM2,EGM} in numerical
approach using finite cut-off a la Lepage\cite{lepage} without
modifying Weinberg's power counting; (2) 'Perturbative'
treatments\cite{KSW1,KSW2,vK,BBSvK,NTvK,LvK,PV} with modified power
counting rules. Discussions of various approaches could be found in
the review articles\cite{BBP,BvK,epelrev,EMRMP,
EM_RP503,EM12,Epel13}. The open status of this issue suggests that
we are still elusive of some intricate structures of the chiral
effective theory for nuclear forces. So, it is worthwhile to do
further studies about the structures of the pionfull theory.

Theoretically, it is easy to handle the pionless theory where pions
are integrated out and expanded into the contact interactions.
Previously, it has been studied without direct reference to the
pionfull one as its adjacent 'underlying' theory. (The
renormalization of this theory could be readily settled using a
'perturbative' scheme based on modified power counting
rules\cite{KSW1,KSW2}. It is also tractable within nonperturbative
regime thanks to the trick of Ref.\cite{PBC} with a general
parametrization of divergences\cite{PRC71,epl85,epl94,5537,JPA42}.)
Then, it is natural to inquire about the detailed mapping between
the pionfull and pionless theories. Through such studies, we may be
able to trace the intricacy of the pionfull theory for nuclear
forces. Therefore, from this report on, we will compute and analyze
the mapping or matching between pionfull and pionless theories for
nuclear forces. The relativistic and non-relativistic formulations
of the standard chiral effective theory\cite{EM_RP503} will be
employed in an interactive manner, which could help to fully reveal
the intricate structures of the pionfull theory, especially to sort
out the subtle issues involving loop integrations. Complementary to
the nonperturbative approaches\cite{EGKM,low-k V,Naka}, our analysis
will be performed at the level of diagrams.

Meanwhile, it is also interesting to see what kind of pionless
theory could be resulted from the various modified power counting
schemes of pionfull theory. For example, we will also compute with
the prescription recently proposed by BKV\cite{BKV}. Recently,
basing on analysis using closed-form $T$ matrices, we found it is
favorable to proceed with an EFT scenario with conventional power
counting\cite{epl85,epl94,5537}. So, it is interesting to see which
or what scenario could be justified from mapping analysis.
Furthermore, it is also interesting to see how various prescription
parameters in pionless theory arise from the pionfull theory via
matching, a more challenging task to be pursued in future. The
mapping perspective may also be valuable for many physical issues
that facilitate EFT descriptions, especially non-relativistic EFT's
with nonperturbative divergences and/or infrared (IR) enhancement
from pinching poles.

This report is organized as follows: The pionfull and pionless
Lagrangians in use are given in Sec. II. In Sec. III, we calculate
the leading contact coupling induced from loop diagrams in pionfull
theory. Sec. IV will be devoted to some general discussions about
our results, where the mapping using BKV prescription will also be
calculated and discussed. The summary will be given in Sec. V.
\section{EFT's for $NN$ scattering}
\subsection{Pionfull EFT}The relativistic Lagrangian we will use
reads (following the notations of Ref.\cite{EM_RP503})\bea\mathcal
{L}_{EFT(\pi)}&=&\mathcal{L}_{\pi\pi}+\mathcal{L}_{\pi{N}}+\mathcal
{L}_{NN}+\cdots,\\\mathcal{L}_{\pi\pi}&=&\frac{1}{2}\partial_\mu\bm
{\pi}\cdot\partial^\mu\bm{\pi}-\frac{1}{2}m^2_\pi\bm{\pi}^2+
\mathcal{O}\left(\bm{\pi}^4\right),\\\mathcal{L}_{\pi{N}}&=&\bar
{\Psi}\left[i\gamma^\mu\partial_\mu-M_N-\frac{g_{A}}{2f_\pi}\gamma^
\mu\gamma^5\bm{\tau}\cdot\partial_\mu\bm{\pi}-\frac{1}{4f^2_\pi}
\gamma^\mu\bm{\tau}\cdot(\bm{\pi}\times\partial_\mu\bm{\pi})+
\mathcal{O}\left(\bm{\pi}^3\right)\right]\Psi,\\\label{NNpionfull}
\mathcal{L}_{NN}&=&-\left(\bar{\Psi}\Gamma_\alpha\Psi\right)\left(
\bar{\Psi}\Gamma^\alpha\Psi\right),\eea with $\Gamma_\alpha$ being
matrices constrained by Lorentz and isospin
invariance\cite{Weinberg}. In non-relativistic formulation where
transparent EFT power counting is feasible, the Lagrangian reduces
to the following form using heavy baryon formalism\bea\mathcal{L}_
{\pi{N}}&=&\bar{N}\left[i\partial_0+\frac{\nabla^2}{2M_N}-\frac{g_
{A}}{2f_\pi}\bm{\tau}\cdot(\bm{\sigma}\cdot\nabla)\bm{\pi}-\frac{1}
{4f^2_\pi}\bm{\tau}\cdot(\bm{\pi}\times\partial_0\bm{\pi})+\mathcal
{O}(\bm{\pi}^3)\right]N,\\\label{NNHBpionfull}\mathcal{L}_{NN}&=&-
\frac{1}{2}C_0\left(\bar{N}N\right)^2+\cdots.\eea Here the contact
couplings should assume the contributions from heavy mesons, etc.,
and scale as:\bea\label{C_0pionfull}C_0\sim\frac{4\pi}{M_N\Lambda_{
(\pi)}},\ \cdots \quad(\Lambda_{(\pi)}\sim4,5m_\pi)\eea with
$\Lambda_{(\pi)}$ being the upper scale of the pionfull EFT.
\subsection{Pionless EFT}After integrating out pions and the
processes above the scale of pion mass, one could further arrive at
a simpler effective theory with only non-relativistic nucleon
degrees and contact interactions among them:\bea\mathcal{L}_{EFT(
\not\pi)}=\bar{N}\left(i\partial_0+\frac{\nabla^2}{2M_N}\right)N-
\frac{1}{2}C^{(\not\pi)}_0\left(\bar{N}N\right)^2+\cdots,\eea with
$\cdots$ representing other contact interactions. Now these the
contact couplings in pionless theory have incorporated contributions
from the pion-exchange diagrams in pionfull theory,\bea\label
{pionlessC_0}{C}^{(\not\pi)}_0=C_0+\hat{T}^{(\pi)}_{NN}(\bm{0},\bm{
0}),\ \cdots,\eea where the counterterms for renormalizing the loop
integrals in $\hat{T}^{(\pi)}_{NN}(\bm{0},\bm{0})$ are obviously
provided by the contact coupling $C_0$ defined in the pionfull
theory (c.f. Eq.(\ref{NNHBpionfull})). As pions are the lightest
quanta for mediating strong forces between nucleons, it is natural
to anticipate that pion-exchange diagrams should dominate the
contributions to the pionless contact couplings, e.g.,\bea\label
{C_0pionless}C^{(\not\pi)}_0\sim\frac{4\pi}{M_N\Lambda_{(\not\pi)}
},\ \cdots\quad(\Lambda_{(\not\pi)}\sim m_\pi).\eea Below, we will
study such contributions, which may shed some light on the intricate
structures of the pionfull theory for nuclear forces.
\section{Mapping into pionless EFT}In the pionfull theory, the $NN$
scattering diagrams could be classified into $2N$-irreducible and
$2N$-reducible ones, which are viewed as pion-exchange $NN$
potential and scattering amplitudes, respectively.
\subsection{2$N$-irreducible diagrams with pions}The
$2N$-irreducible diagrams in pionfull EFT have been computed up to
next-to-next-to-next-to-leading order in literature, see
Refs.\cite{kaiser,EGM,empot1,EM1,EM2,EM_RP503}. For our purpose
below, it suffices to demonstrate with the one-pion exchange (OPE)
and two-pion exchange (TPE) components
($\mathcal{O}(Q^2)$)\cite{kaiser}:\bea{V}_{1\pi}(\bm{q})&=&-\frac{g
_A^2}{4f^2_\pi}\bm{\tau}_1\cdot\bm{\tau}_2\frac{\bm{\sigma}_1\cdot
\bm{q}\ \bm{\sigma}_2\cdot\bm{q}}{q^2+m^2_\pi},\\V_{2\pi}(\bm{q})&=
&\bm{\tau}_1\cdot\bm{\tau}_2W_C+\bm{\sigma}_1\cdot\bm{\sigma}_2V_S
+\bm{\sigma}_1\cdot\bm{q}\ \bm{\sigma}_2\cdot\bm{q}V_T,\\\label
{Wc-TPE}W_C&=&\frac{-1}{384\pi^2f^4_\pi}\left\{\left[4m^2_\pi\left
(5g^4_A-4g^2_A-1\right)+q^2\left(23g^4_A-10g^2_A-1\right)+\frac{48
g^4_Am^4_\pi}{4m_\pi^2+q^2}\right]L(q)\right.\nonumber\\&&+\left[6
m^2_\pi\left(15g^4_A-6g^2_A-1\right)+q^2\left(23g^4_A-10g^2_A-1
\right)\right]\ln\frac{m_\pi}{\mu}\nonumber\\&&\left.+4m^2_\pi\left
(4g^4_A+g^2_A+1\right)+\frac{q^2}{6}\left(5g^4_A-26g^2_A+5\right)
\right\},\\V_T&=&-\frac{1}{q^2}V_S=-\frac{3g^4_A}{64\pi^2f^4_\pi}
L(q),\eea where\bea{L}(q)\equiv\frac{\sqrt{4m^2_\pi+q^2}}{q}\ln
\frac{\sqrt{4m^2_\pi+q^2}+q}{2m_\pi},\quad q\equiv|\bm{q}|,\quad
\bm{q}\equiv\bm{p}-\bm{p}^\prime,\eea with $\bm{p},\bm{p}^\prime$
being the external momenta for a nucleon. Below, the
renormalization-scale-dependent terms $(\propto\ln\frac{m_\pi}
{\mu})$ will be discarded (by putting $\mu=m_\pi$) as in
Refs.\cite{EGM,empot1,EM1,EM2}, as the qualitative status would
remain the same. Besides this, the $W_C$ of TPE given in
Ref.\cite{EGM} only contains the term in the first line of
Eq.(\ref{Wc-TPE}).

Now, we perform the low-energy expansion to extract contributions to
the contact couplings in pionless EFT. We focus on $C_0$ (the
superscript '$(\not\!\!\pi)$' will be dropped henceforth), to which
OPE contributes nothing due to the derivative $\pi N$ coupling!
While the TPE's contribution differs a little across literature
(below, the superscripts '$^{(\text{\tiny KBW})}$' and
'$^{(\text{\tiny EGM})}$' refer to Ref.\cite{kaiser} and
Ref.\cite{EGM}, respectively):\bea\label{C_KBW}V^{(\text{\tiny
KBW})}_{2\pi}\Rightarrow&&C^{(\text{\tiny KBW})}_{0\tau}=-\frac
{g^4_Am^2_\pi}{8\pi^2f^4_\pi},\quad\Lambda^{(\text{\tiny KBW})}_
{(\not\pi,\tau)}\equiv-\frac{4\pi}{M_NC^{(\text{\tiny KBW})}_{0
\tau}}=\frac{32\pi^3f^4_\pi}{g^4_AM_Nm_\pi^2},\\\label{C_EGM}V^
{(\text{\tiny EGM})}_{2\pi}\Rightarrow&&C^{(\text{\tiny EGM})}_{0
\tau}=-\frac{g^4_Am^2_\pi}{12\pi^2f^4_\pi},\quad\Lambda^{(\text
{\tiny EGM})}_{(\not\pi,\tau)}\equiv-\frac{4\pi}{M_NC^{(\text{\tiny
EGM})}_{0\tau}}=\frac{48\pi^3f^4_\pi}{g^4_AM_Nm_\pi^2}, \eea with
the scale $\Lambda$ thus extracted being of order $10^3$ MeV (see
Table 2), much larger than the upper scale of pionless EFT that is
of order $m_\pi$. (We only extracted the terms of order $g^4_A$ as
$g_A>1.2$ and including the terms of lower $g_A$ power would not
alter the magnitude order of our results.) In the pionfull theory,
the constants given in Eqs.(\ref{C_KBW},\ref{C_EGM}) are the leading
contribution to pionless $C_0$ from $2N$-irreducible diagrams
(potential).

Comparing with power counting in Eq.(\ref{C_0pionless}), such
contributions are too small. That means, the dominant contribution
to the pionless $C_0$ could not come from such $2N$-irreducible
diagrams. Then we are left with the diagrams containing iterations
of pion-exchange potential, i.e., the $2N$-reducible diagrams. The
simplest case is the once-iterated OPE diagram, which has been
computed long ago by the Munich group\cite{kaiser}. Below, we will
reanalyze it from the mapping perspective through an 'interactive'
use of non-relativistic and relativistic formulations. The
calculations will be done using conventional chiral lagrangian and
regularization schemes without additional prescriptions like
PDS\cite{KSW1} or intermediate manipulations like IR
regularization\cite{becher1,becher2}.
\subsection{$2N$-reducible diagrams with pions: 3-dimensional
non-relativistic calculation} Our parametrization below is based on
Ref.\cite{EM_RP503}. In non-relativistic formulation, the
once-iterated OPE diagram reads\bea{T}^{(it)}_{1\pi}(\bm{p},\bm{p}^
\prime)=\frac{g^4_A}{16f^4_\pi}(3-2\bm{\tau}_1\cdot\bm{\tau}_2)\int
\frac{d^3\bm{l}}{(2\pi)^3}\frac{\bm{\sigma}_1\cdot\bm{q}_1\ \bm{
\sigma}_2\cdot\bm{q}_1\ \bm{\sigma}_1\cdot\bm{q}_2\ \bm{\sigma}_2
\cdot\bm{q}_2}{\left(\bm{q}_1^2+m^2_\pi\right)\left(\bm{q}_2^2+m^2_
\pi\right)\left(E_{N;p}-\frac{\bm{l}^2}{M_N}+i\epsilon\right)},\eea
with $\bm{q}_1=\bm{p}+\bm{l},\ \bm{q}_2=\bm{p}^\prime+\bm{l},\ {E}_
{N;p}\equiv\sqrt{\bm{p}^2+M^2_N}.$ Here the superscript "$({it})$"
indicates the once-iterated OPE diagram.

To extract the contribution to $C_0$, we compute the following\bea
&&{T}^{(it)}_{1\pi}(\bm{0},\bm{0})=-\frac{g^4_AM_N}{16f^4_\pi}(3-2
\bm{\tau}_1\cdot\bm{\tau}_2)I_4(\bm{0}),\\\label{I40def}&&{I}_4(\bm
{0})\equiv\int\frac{d^3\bm{l}}{(2\pi)^3}\frac{\bm{l}^2}{E^4_{\pi;l}
},\eea with ${E}_{\pi;l}\equiv\sqrt{\bm{l}^2+m^2_\pi}$. In standard
dimensional and cutoff schemes, we have\bea\label{I40}I_4(\bm{0})=
\left\{\begin{array}{l}\displaystyle-\frac{3m_\pi}{8\pi},\
($dimensional$)\\\\\displaystyle-\frac{3m_\pi}{8\pi}+\frac{\Lambda}
{2\pi^2},\ ($cutoff$)\end{array}\right.\eea As will be seen in
Sec.3.3, the linear divergence here is an artifact introduced by
non-relativistic approximation. So, we take that\bea\label{Tit00}T^
{(it)}_{1\pi}(\bm{0},\bm{0})=\frac{3g^4_AM_Nm_\pi}{128\pi{f}^4_\pi}
(3-2\bm{\tau}_1\cdot\bm{\tau}_2).\eea This is essentially what the
once-iterated OPE diagram contributes to the leading coupling $C_0$
in pionless theory, the contribution to $I_4(\bm{0})$ from pionless
region is negligible:\bea{I}_4^{(\not\pi)}(\bm{0})\equiv\int_{\leq
m_\pi}\frac{d^3\bm{l}}{(2\pi)^3}\frac{\bm{l}^2}{E^4_{\pi;l}}=-
\varepsilon_4^{(\not\pi)}\frac{3m_\pi}{8\pi},\quad\left|\varepsilon
_4^{(\not\pi)}\right|=\frac{10-3\pi}{6\pi}\approx3.05\times10^{-2}
\ll1.\eea Obviously, the suppression of the contribution from
pionless range is due to the derivative pion-nucleon coupling.

To be more accurate, one may exclude this 3 percent in identifying
the dominant contribution to $C_0$:\bea&&C^{(it)}_0+C^{(it)}_{0\tau}
\bm{\tau}_1\cdot\bm{\tau}_2\equiv{T}^{(it)}_{1\pi}(\bm{0},\bm{0})
\left(1-\varepsilon_4^{(\not\pi)}\right)\\&&{C}^{(it)}_0=\frac{9g^4
_AM_Nm_\pi}{128\pi{f}^4_\pi}\left(1-\varepsilon_4^{(\not\pi)}\right
),\quad{C}^{(it)}_{0\tau}=-\frac{3g^4_AM_Nm_\pi}{64\pi{f}^4_\pi}
\left(1-\varepsilon_4^{(\not\pi)}\right).\eea Following the standard
parametrization: $C_0=\pm{4\pi}{M^{-1}_N\Lambda^{-1}_{(\not\pi)}}$,
we have\bea\Lambda^{({it})}_{(\not\pi)}=\frac{512\pi^2f^4_\pi}{9g^
4_AM^2_Nm_\pi\left(1-\varepsilon_4^{(\not\pi)}\right)},\quad\Lambda
^{({it})}_{(\not\pi,\tau)}=\frac{256\pi^2f^4_\pi}{3g^4_AM^2_Nm_\pi
\left(1-\varepsilon_4^{(\not\pi)}\right)},\eea which is of the order
of pion mass provided the popular choices for $M_N$, $m_\pi$,
$f_\pi$ and $g_A$ are made. In table 1 and table 2, the 3 percent
deduction is not included as it could not affect our conclusions.
\subsection{$2N$-reducible diagrams with pions: 4-dimensional
relativistic calculation} 
\begin{figure}[h]\begin{center}
\hspace*{-0.5cm}\vspace*{-0.25cm}\resizebox{13.5cm}{!}
{\includegraphics{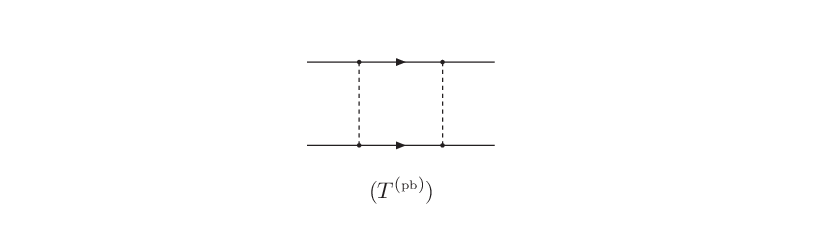}}\caption{\small Planar box
diagram in relativistic formulation}\label{Tpb}\end{center}
\end{figure}

In relativistic formulation, the
once-iterated OPE diagram is contained in the following planar box
diagram (Fig.\ref{Tpb}):\bea{T}^{(\texttt{\tiny pb})}(\bm{p},\bm{p}
^\prime)&=&\frac{g^4_A}{16f^4_\pi}\int\frac{d^4l}{(2\pi)^4}\frac{1}
{\left(q_1^2-m^2_\pi\right)\left(q_2^2-m^2_\pi\right)}\bar{u}_1(
\bm{p}^\prime)(-\not\!\!q_2)\gamma^5\tau_1^b\frac{1}{\not\!k-M_N}
\not\!\!q_1\gamma^5\tau_1^au_1(\bm{p})\nonumber\\&&\times\bar{u}_2
(-\bm{p}^\prime)\not\!\!q_2\gamma^5\tau_2^b\frac{1}{\not\!k^\prime
-M_N}(-\not\!\!q_1)\gamma^5\tau_2^au_2(-\bm{p})\eea with momentum
flows chosen as in Ref.\cite{EM_RP503}: ${q}_1=(l^0,\bm{p}-\bm{l}),\
{q}_ 2=(l^0,\bm{p}^\prime-\bm{l}),\ {k}=(E_{N;p}-l^0,\bm{l}),\ {k}^
\prime=(E_{N;p}+l^0,-\bm{l}).$

Again, we are interested in the situation when external momenta are
zero. After some works (see Appendix A), we have\bea\label{4dT-pb}T
^{(\texttt{\tiny pb})}(\bm{0},\bm{0})&=&-\frac{g^4_A}{128\pi^2f^4_
\pi}(3-2\bm{\tau}_1\cdot\bm{\tau}_2)\left[\alpha_NM^2_N+\alpha_{N
\pi}M_Nm_\pi+\alpha_{\pi}m^2_\pi\right],\eea where\bea\alpha_N&
\equiv&\Gamma(\epsilon)+3-\ell_{N},\quad\alpha_{N\pi}\equiv\frac{(2
-6\varrho)\arctan\sqrt{4\varrho-1}}{\varrho\sqrt{1-(4\varrho)^{-1}}
},\nonumber\\\alpha_\pi&\equiv&-\frac{\Gamma(\epsilon)+1-\ell_{\pi}}
{4}-3+\frac{8\varrho-3}{2\varrho}\ln\varrho+\frac{(10\varrho-3)
\arctan\sqrt{4\varrho-1}}{\varrho\sqrt{4\varrho-1}}.\eea Obviously,
'$\alpha_\pi{m}^2_\pi$' is what one would formally expect for a
standard TPE component of $NN$ potential. '$\alpha_{N\pi}M_Nm_\pi$'
is a definite (or nonlocal) term that comes from $I_{2+-}$, it is
just the dominant contribution from the box diagram to the pionless
$C_0$, see below. According to Eq.(\ref{pionlessC_0}), the
logarithmic divergences in $T^{(\texttt{\tiny pb})}(\bm{0},\bm{0})$
could be subtracted by the counterterm from $C_0$ in the pionfull
theory.

However, there are also 'offensively' large, finite and local items
in $\alpha_N$ and $\alpha_\pi$ that obviously violate the power
counting of chiral effective theory. To resolve this problem, we
note that the pionfull theory actually lives in non-relativistic
regime as $\Lambda_{(\pi)}$ lies well below $M_N$. Then, after
contour integration, $M_N$ activates a division of loop momentum
space into low/non-relativistic and high/relativistic regions: In
the low region, there are at most chiral divergences as all momenta
and $m_\pi$ could be treated as small scales; While in the high
region, only external momenta and $m_\pi$ are smaller scales that
facilitate expansions, resulting in local operators of low-energy
degrees with 'offensively' large coefficients that should be removed
in order to stay in non-relativistic regime. Therefore, the
counterterms in the relativistic formulation should contain two
components: one removes the 'offensively' large relativistic
contributions, another subtracts the chiral divergences. The first
component is just the necessary tool required by the decoupling
theorem\cite{decouplth1,decouplth2} underling effective field
theories.

Let us illustrate with the definite integral $I_{2+-}$ that
interests us most. To enter non-relativistic regime, one first picks
up the low-lying poles at $E_{N;l}-M_N\approx\frac{\bm{l}^2}{2M_N}$
(nucleon) and $E_{\pi;l}$ (pion) in contour integration and then
expand the resultants in terms of $1/M_N$ in the low region. For
$I_{2+-}$, we have:\bea\label{NR}&&\left.I_{2+-}\right|_{NR}\equiv
\left.\int\frac{d^3\bm{l}}{(2\pi)^3}\left(\oint\frac{dl_0}{2\pi}
\frac{l^2_0}{A_\pi{A}_+A_-}\right)\right|_{NR}=\frac{i(4M_NI_{N}+I_
{\pi})}{64M^4_N},\\&&I_{N}\equiv\int\frac{d^3\bm{l}}{(2\pi)^3}\frac
{\bm{l}^2}{E_{\pi;l}^4}=I_4(\bm{0}),\quad{I}_{\pi}\equiv\int\frac{d
^3\bm{l}}{(2\pi)^3}\frac{m^4_\pi+4m^2_\pi\bm{l}^2-4M^2_NE^2_{\pi;l}
}{E_{\pi;l}^{5}},\eea with $I_N$ and $I_\pi$ denoting the outcomes
from the low-lying nucleon and pion poles, respectively. From
Eqs.(\ref{I40def},\ref{I40},\ref{I2+-},\ref{NR}), we see that, $I_4
(\bm{0})$ actually comes from the following nonlocal piece in the
definite integral $I_{2+-}$ (see Appendix A):\bea\frac{1}{64\pi^2M^
2_N}\times\frac{(2-6\varrho)\arctan\sqrt{4\varrho-1}}{\varrho\sqrt
{4\varrho-1}}=\frac{1}{16M^3_N}\left\{-\frac{3m_\pi}{8\pi}\left[1+
o\left(\varrho^{-\frac{1}{2}}\right)\right]\right\}.\eea Obviously,
the linear divergence in $I_4(\bm{0})$ is 'generated' with the
non-relativistic truncation of a definite integral in relativistic
formulation, justifying our choice in Sec. III.B. In the meantime,
the following terms are subtracted:\bea\delta{I}_{2+-}=I_{2+-}-
\left.I_{2+-}\right|_{NR}=\frac{i}{(8\pi)^2}\frac{1}{M^2_N}\left\{
\left[\Gamma(\epsilon)-\ell_{N}\right]\left(1-\varrho^{-1}\right)+2
+2\varrho^{-1}+o\left(\varrho^{-\frac{3}{2}}\right)\right\},\eea
which are just the outcomes of the nucleon poles at $E_N\pm M_N$
integrated over the high region and other relativistic corrections.
Collecting these 'subtracted' terms for $T^{(\texttt{\tiny pb})}
(\bm{0},\bm{0})$, we have the following counterterm\bea-\check{
\Delta}T^{(\texttt{\tiny pb})}(\bm{0},\bm{0})=\frac{g^4_A(3-2\bm
{\tau}_1\cdot\bm{\tau}_2)}{128\pi^2f^4_\pi}\left\{\left(\Gamma(
\epsilon)-\ell_{N}\right)\left(M^2_N-4m^2_\pi\right)+3M^2_N+o\left(
\varrho^{-1}M^2_N\right)\right\},\eea which obviously contains the
'offensively' large items mentioned above, implementing the
'decoupling' of high region contributions. Now it is clear that the
subtraction of the 'offensively 'large terms is an inherent part of
working in non-relativistic regime, an interesting fact lending
itself to the understanding of the proposal for preserving the
conventional power counting in relativistic baryon
$\chi$PT\cite{becher1,becher2,Gege-Scherer}.

Therefore, in non-relativistic regime, the box diagram decomposes
into $2N$-reducible and $2N$-irreducible components as below:\bea
\left.{T}^{(\texttt{\tiny pb})}(\bm{0},\bm{0})\right|_{NR}={T}^{(
\texttt{\tiny pb})}(\bm{0},\bm{0})-\check{\Delta}T^{(\texttt{\tiny
pb})}(\bm{0},\bm{0})=T^{(it)}_{1\pi}(\bm{0},\bm{0})+V^{(\texttt{
\tiny pb})}_{2\pi}(\bm{0}),\eea with\bea{V}^{(\texttt{\tiny pb})}_
{2\pi}(\bm{0})=\frac{g^4_A}{128\pi^2f^4_\pi}(3-2\bm{\tau}_1\cdot\bm
{\tau}_2)m^2_\pi\left\{4-\frac{15}{4}\left[\Gamma(\epsilon)+1-\ell_
\pi\right]\right\}\eea being the (bare) $2N$-irreducible component:
part of the TPE potential\cite{kaiser,EM_RP503} as the crossed box
diagram is not included here. Obviously, ${V}^{(\texttt{\tiny pb})}
_{2\pi}(\bm{0})$ is the outcome of the pion pole while $T^{(it)}_{1
\pi}(\bm{0},\bm{0})$ is the outcome of the low-lying nucleon pole.
The divergence in $V^{(\texttt{\tiny pb})}_{2\pi}$ is chiral and
could be subtracted using the following chiral counterterm\bea
\delta{V}^{(\texttt{\tiny pb})}_{2\pi}(\bm{0})=\frac{15g^4_A}{512
\pi^2f^4_\pi}(3-2\bm{\tau}_1\cdot\bm{\tau}_2)m^2_\pi\left[\Gamma(
\epsilon)+1-\ell_\pi\right].\eea

Now we arrive at the finite contributions to the pionless coupling
$C_0$ from the planar box diagram that also decompose into two
components\bea&&{C}^{(it)}_0+{C}^{(it)}_{0\tau}\bm{\tau}_1\cdot\bm{
\tau}_2\equiv{T}^{(it)}_{1\pi}(\bm{0},\bm{0})=\frac{3g^4_AM_Nm_\pi}
{128\pi f^4_\pi}(3-2\bm{\tau}_1\cdot\bm{\tau}_2),\\&&C^{(irr)}_0+C^
{(irr)}_{0\tau}\bm{\tau}_1\cdot\bm{\tau}_2\equiv{V}^{(\texttt{\tiny
pb})}_{2\pi;R}(\bm{0})=\frac{g^4_Am^2_\pi}{32\pi^2f^4_\pi}(3-2\bm{
\tau}_1\cdot\bm{\tau}_2),\eea with the ratio\footnote{The
prescription dependence of ${V}^{(\texttt{\tiny pb})}_{2\pi;R}$
should not affect this ratio materially.}\bea\frac{C^{(it)}_0}{C^{
(irr)}_0}=\frac{3\pi}{4}\varrho^{\frac{1}{2}}=\frac{3\pi M_N}{4m_
\pi}\approx16.03\gg1\eea demonstrates clearly the dominance of the
$2N$-reducible component within planar box diagram. In relativistic
formulation, there would be small relativistic corrections that will
not alter this dominance. The crossed box diagram contains no
contribution to $C_0$ except a $2N$-irreducible piece that belongs
to TPE\cite{EM_RP503}.

Here, some remarks are in order: (1) Concerning the contributions to
the coupling $C_0$ in pionless EFT, a definite and hence nonlocal
item from the $2N$-reducible component of the box diagram is a
dominant in comparison with that from $2N$-irreducible diagrams or
components. The same might also happen to higher pionless couplings.
(2) Exploiting the virtues of both non-relativistic and relativistic
formulations, we identified the rationale for subtracting
'offensively' large terms in the pionfull theory for nuclear forces.
Recently, the virtue that relativistic formulation embodies
less UV divergences has also been exploited in Ref\cite{EG1207},
resulting in a modified Weinberg approach for nuclear forces where
former pathologies could be removed or diminished. (3) The following
strategy surfaces in our analysis: a) In relativistic form, the
'offensively' large contributions from high region should be
subtracted to stay in non-relativistic regime, the rest divergences
are chiral ones and tractable within chiral effective theory; b)
In non-relativistic form, the power divergences in the $2N$-reducible
diagrams are artefact of non-relativistic truncation and could be
treated with dimensional regularization, the $2N$-irreducible ones
are also tractable within chiral effective theory.

The various contributions to the pionless $C_{0\tau}$ are summarized
in table 1 and table 2. In table 2, we also listed the scale
extracted for the isospin-independent coupling $C^{(it)}_0$ in the
last column.\begin{table}[ph]\label{various-C_0}\caption{Various
contributions to $C_{0\tau}$ and $\Lambda_{(\not\pi,\tau)}$}
\begin{center}\begin{tabular}{c|cccc}\hline\hline&$\quad\quad$OPE$
\quad$&$\quad$TPE(KBW)$\quad$&$\quad$TPE(EGM)$\quad$&$\quad$
ITERATION$_\tau\quad$\\\hline\\$C_{0\tau}$&$\quad0$&$\displaystyle-
\frac{g^4_Am^2_\pi}{8\pi^2f^4_\pi}$&$-\displaystyle\frac{g^4_Am^2_
\pi}{12\pi^2f^4_\pi}$&$\quad-\displaystyle\frac{3g^4_AM_Nm_\pi}{64
\pi f^4_\pi}$\\&&&&\\$\Lambda_{(\not\pi,\tau)}$&$\quad\infty$&$
\displaystyle\frac{32\pi^3f^4_\pi}{g^4_AM_Nm_\pi^2}$&$\displaystyle
\frac{48\pi^3f^4_\pi}{g^4_AM_Nm_\pi^2}$&$\quad\displaystyle\frac
{256\pi^2f^4_\pi}{3g^4_AM_N^2m_\pi}$\\&&&&\\\hline\hline
\end{tabular}\end{center}\label{C0}\end{table}
\begin{table}[ph]\label{various-Lambda}
\caption{$\Lambda_{(\not\pi,\tau)}$ and $\Lambda_{(\not\pi)}$ in MeV
with $(f_\pi,m_\pi,M_N)=(92.4,138,939)$ MeV.}
\begin{center}\begin{tabular}{c|cccc}\hline\hline$\quad{g}_A\quad$&$
\quad$TPE(KBW)$\quad$&$\quad$TPE(EGM)$\quad$&$\quad$ITERATION$_{
\tau}$$\ $&$\ $ITERATION$\quad$\\\hline1.26&1604.65&2406.98&200.18&
133.45\\&($\sim11.63m_\pi$)&($\sim17.44m_\pi$)&($\sim1.45m_\pi$)&($
\sim0.97m_\pi$)\\\hline1.29&1460.51&2190.77&182.20&121.46\\&($\sim
10.58m_\pi$)&($\sim15.88m_\pi$)&($\sim1.32m_\pi$)&($\sim0.88m_\pi$
)\\\hline1.32&1332.20&1998.29&166.19&110.79\\&($\sim9.65m_\pi$)&($
\sim14.48m_\pi$)&($\sim1.20m_\pi$)&($\sim0.80m_\pi$)\\\hline\hline
\end{tabular}\end{center}\label{scaleinC0}\end{table}
\section{Region division, enhancement and mapping}
\subsection{General reasoning}In relativistic formulation of any
EFT, loop momentum scale extends to infinity. However, the vast
region above the upper scale of EFT, $[\Lambda_{(\texttt{\tiny
EFT})},\infty)$, is actually superfluous. For theories with light
mass scales, the vast superfluous region is of no harm. Things
become complicated when an EFT actually lives in non-relativistic
regime: Offensively large terms have to be subtracted to stay in
non-relativistic regime, then intricacies arise due to the infrared
enhancement in non-relativistic regime. In the pionfull theory for
nuclear forces, the pions mass facilitates a further division of the
low region '$U_{(\pi)}$' into pionless region '$U_{(\not\pi)}$' and
its complement '$\tilde{U}_{(\pi)}$':\bea{U}_{(\pi)}={U}_{(\not\pi)
}\cup\tilde{U}_{(\pi)},\quad{U}_{(\not\pi)}\equiv\left[0,\Lambda_{(
\not\pi)}\right),\quad\tilde{U}_{(\pi)}\equiv\left[\Lambda_{(\not
\pi)},\Lambda_{(\pi)}\right).\eea Intricacies actually lie in
$\tilde{U}_{(\pi)}$, where low-lying nucleon poles dominate the
contributions to pionless couplings due to infrared
enhancement\footnote{Literally, as high and low regions are
separated by nucleon mass $M_N$, an extra region $\delta{U}_{low}=
[\Lambda_{(\pi)},M_N)$ is implicitly included in the loop
integration. We are not clear yet the roles played by this extra
region.}. Of course, it remains to see how higher diagrams behave in
this region, especially how the low-lying nucleon poles in these
diagrams contribute to pionless couplings!

Technically, the dominance of iterated OPE in the contributions to
pionless $C_0$ is primarily due to the dominance of $4M_NI_N$ over
$I_\pi$, as the low-lying nucleon poles tend to pinch in the $2N$
reducible component of the box diagram. We also need that the
contribution from the pionless region to $I_N$ is negligible, which
is guaranteed by the derivative $\pi N$ coupling. Thus, for the
anomalous dominance of $2N$ reducible diagrams, we need: (1)
non-relativistic regime where the low-lying nucleon poles tend to
pinch; (2) derivative $\pi N$ coupling to suppress the contributions
from pionless region so that the pionfull region $\tilde{U}_{(\pi)}$
holds the bulk contributions, (3) clear separation of mass scales to
make the enhancement materialize, i.e., $\sqrt{\varrho}\gg1$. Of
course, the 'offensively' large relativistic components must be
entirely excluded or subtracted in the first place. Otherwise, the
whole theory will be overwhelmed by the high region, which is
totally unacceptable. Unless profound changes are made to
substantially invalidate the above three features or conditions, the
dominance of iterated diagrams is doomed to happen, right within the
region $\tilde{U}_{(\pi)}$\footnote{This is also reflected by the
fact that the scale $\sqrt{M_Nm_\pi}$ from infrared enhancement is
close to half of $\Lambda_{(\pi)}$:$\sqrt{M_Nm_\pi}\sim2.61m_\pi
\sim\frac{\Lambda_{(\pi)}}{2}.$}. So, in this perspective, the real
issue of the pionfull theory for nuclear forces stems from the low
region $\tilde{U}_{(\pi)}$, not from the high region whose bulk
contributions must be subtracted according to decoupling theorem to
stay in non-relativistic regimes. Hence, one should either work
entirely in nonperturbative regime for OPE or alter the organization
of the theory right within the region $\tilde{U}_{(\pi)}$. Actually,
the scale $\sqrt{M_Nm_\pi}$ was also shown to be associated with
radiative pions and slow down the convergence of chiral expansion of
$NN$ forces\cite{Mondejar}.

In addition, our primary analysis using once-iterated OPE diagram
seems to support the conventional power counting rules for pionless
EFT given in Sec. 2.2. Of course, extensive studies about more
diagrams are needed. Below, we wish to see what could happen to the
pionfull-pionless mapping in BKV prescription\cite{BKV}.
\subsection{Mapping in BKV prescription}In BKV prescription, the
higher modes are separated out from OPE using the following
means\cite{BKV}:\bea{V}_{1\pi}^{(\texttt{\tiny BKV})}(\bm{q})=-
\frac{g_A^2}{4f^2_\pi}\bm{\tau}_1\cdot\bm{\tau}_2\left[\bm{\sigma}
_1\cdot\bm{q}\ \bm{\sigma}_2\cdot\bm{q}\left(\frac{1}{q^2+m^2_\pi}-
\frac{1}{q^2+\lambda_{\texttt{\tiny BKV}}^2}\right)+\frac{\lambda_
{\texttt{\tiny BKV}}^2}{q^2+\lambda_{\texttt{\tiny BKV}}^2}\right],
\eea with $\lambda_{\texttt{\tiny BKV}}$ (set at 750 MeV) being the
separation scale. Obviously, this 'OPE' contributes to the leading
coupling:\bea{C}^{(\texttt{\tiny BKV})}_{0\tau}\bm{\tau}_1\cdot\bm
{\tau}_2=V^{(\texttt{\tiny BKV})}_{1\pi}(\bm{0})=-\frac{g_A^2}{4f^
2_\pi}\bm{\tau}_1\cdot\bm{\tau}_2\eea with '$\frac{4\pi}{M_NC_{0
\tau}}$' being close to $2m_\pi$, so it will be taken as the leading
contribution to $C_{0\tau}$ from pion-exchange potential as higher
order pion-exchange (TPE, etc.) should also be sub-leading in BKV
prescription.

In the meantime, the iteration of this 'OPE' gives ($\theta\equiv
\frac{\lambda_{\texttt{\tiny BKV}}}{m_\pi}$):\bea{T}^{(it;\texttt
{\tiny BKV})}_{1\pi}(\bm{0},\bm{0})&=&-\frac{g^4_AM_Nm_\pi}{16f^4_
\pi}(3-2\bm{\tau}_1\cdot\bm{\tau}_2)\left[I_{4;(\texttt{\tiny BKV})
}(\bm{0})+\bm{\sigma}_1\cdot\bm{\sigma}_2I_{4\bm{\sigma};(\texttt
{\tiny BKV})}(\bm{0})\right]\nonumber\\&=&-\frac{g^4_AM_Nm_\pi}{16f
^4_\pi}(3-2\bm{\tau}_1\cdot\bm{\tau}_2)\left\{\frac{2\theta^2-
\theta+1}{8\pi(1+\theta)}+\frac{\bm{\sigma}_1\cdot\bm{\sigma}_2
{\theta}^2}{6\pi(1+\theta)}\right\},\eea from which we could find
(using $\lambda_{\texttt{\tiny BKV}}=750$ MeV)\bea\left|\frac{C^{
(it;\texttt{\tiny BKV})}_{0\tau}}{C^{(\texttt{\tiny BKV})}_{0\tau}}
\right|=\frac{g^2_AM_Nm_\pi}{16\pi f^2_\pi}\frac{2\theta^2-\theta+1}
{(1+\theta)}\approx2.564g^2_A,\eea which moves from 4.07 to 4.47 as
$g_A$ varies from 1.26 to 1.32. Note that $I_{4;(\texttt{\tiny BKV}
)}$ still mainly comes from the pionfull region as $I_{4;(\texttt
{\tiny BKV})}(<m_\pi)/I_{4;(\texttt{\tiny BKV})}\approx15.6\%$.
Excluding the pionless region, the above ratio becomes $\approx
2.163g^2_A,$ which varies from 3.43 to 3.77 when $g_A$ varies as
above. Thus, the dominance of the iterated 'OPE' still happens,
leading again to anomalous mapping between pionfull and pionless
theories. This is because that the BKV modification is mainly
introduced to tame the UV behavior of OPE in triplet channels, it
does not invalidate the three conditions for dominance of iterated
diagrams.

Here, we note in passing that the scaling of the dominant
contribution to pionless $C_0$ in BKV is approximately $C^{(it;
\texttt{\tiny BKV})}_{0\tau}\sim\frac{4\pi}{M_N\Lambda_{(\not\pi;
\texttt{\tiny BKV})}}$ with $\Lambda_{(\not\pi;\texttt{\tiny BKV})
}\approx\frac{1}{2}m_\pi$, quite a distance from the KSW scaling
$C_{0}\sim\frac{4\pi}{M_NQ}$ with $Q\sim35$MeV, thus the KSW
scaling for pionless EFT is not quite realized yet in the BKV
prescription. Of course, further studies of higher diagrams are
needed for a conclusive judgement.
\subsection{Anomalous mapping and 'dibaryon' like configuration}
\begin{figure}[h]\begin{center}
\hspace*{-0.6cm}\vspace*{-0.25cm}\resizebox{18cm}{!}
{\includegraphics{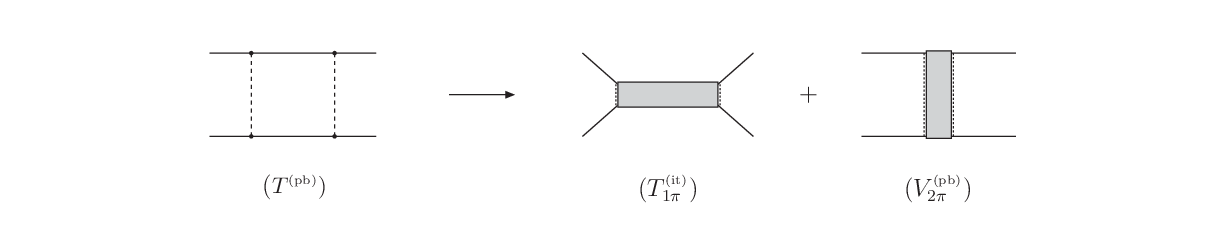}}\caption{\small Spacetime
configurations of $T^{\texttt{\tiny(it)}}_{1\pi}$ and $V^{\texttt
{\tiny(pb)}}_{2\pi}$ from decomposition of $T^{\texttt{\tiny(pb)}}
$}\label{NN}\end{center}\end{figure} At this stage, we may digress a
little about the configuration of the anomalously dominant item that
is nonlocal in the planar box diagram. Due to the hierarchy between
the low energy nucleon pole and pion mass, $E_N\approx\frac{Q^2}
{2M_N}\ll m_\pi$, the dominant item essentially comes from the
contributions of such a spacetime configuration that the two
on-shell nucleons moves 'together' with a small space separation
(separated by potential pion) over a large spacetime distance. That
means, such a configuration is close to that of the propagation of a
composite object made of two potential-pion-exchanging baryons as an
intermediate state, hinting us at a dibaryon like object as an
intermediate state, as illustrated in Fig.\ref{NN}. Of course, at
present stage, it is merely a loose and speculative analogue based
on a simple analysis on the box diagram with vanishing external
momenta of nucleons. Further studies (especially on diagrams of more
pion exchanges with non-vanishing external momenta) are needed for
firmer conclusions. The above analysis reminds us that an
appropriate incorporation of pertinent degrees may also be of
concern in order to arrive at a better organization of the pionfull
EFT for nuclear forces. For incorporating dibaryon degrees
explicitly in $NN$ scattering and related issues, we refer to
Refs.\cite{dibaryon1,dibaryon2,dibaryon3}. Of course, in such
approaches, the pion exchanges' contributions must be organized in a
manner consistent with the presence of dibaryons to avoid double
counting, at least the parts that might be simulated by dibaryon
like degrees must be sophisticatedly 'subtracted' from the
pion-exchange diagrams. This point is natural to see from the
mapping perspective, but seems to have been overlooked in
literature.
\subsection{Emergence of large scattering lengths}Our foregoing
derivations clearly demonstrated that the pionless coupling $C_0$
essentially comes from a definite item coming from the region
$[\Lambda_{(\not\pi)},\Lambda_{(\pi)})$ in the (underlying)
pionfull theory. Therefore, loop integrations in pionless EFT only
make sense over the region $[0,\Lambda_{(\not\pi)})$, as
contributions from $[\Lambda_{(\not\pi)},\Lambda_{(\pi)})$ have
been assumed by the pionless couplings. In this perspective, the
pionless integrals would become definite items in the (underlying)
pionfull theory that mainly collect contributions from the pionless
region $[0,\Lambda_{(\not\pi)})$. For example, in ${T}_0=\frac{C_0}
{1-C_0I_{0;(\not\pi)}}$ generated by the pionless $C_0$, the
pionless integral$${I}_{0;(\not\pi)}\equiv\int\frac{d^3\bm{l}}
{(2\pi)^3}\frac{1}{E-{\bm{l}^2}/{M_N}+i\epsilon}=-J_0-i\frac{M_Np}
{4\pi}\quad\left(p\equiv\sqrt{M_NE}\right)$$ should collect
contributions from $[0,\Lambda_{(\not\pi)})$ in a well-defined
manner in the pionfull theory, hence $J_0\sim\frac{M_N}{4\pi}
\Lambda_{(\not\pi)}$. Then, large scattering lengths in $S$ waves
would 'naturally' emerge provided $C_0\sim-\frac{4\pi}{M_N}\Lambda
^{-1}_{(\not\pi)}$:$$\frac{1}{a}=\Re\left[-\frac{4\pi}{M_NT_0}
\right]_{p=0}=-\frac{4\pi}{M_N}\left(\frac{1}{C_0}+J_0\right)=\pm
o\left(\epsilon^\sigma\Lambda_{(\not\pi)}\right),\ \left(\sigma
\geq1,\ \epsilon\sim4^{-1}\right)$$ which is true even after higher
couplings are included\cite{PRC71,epl85,epl94,5537}. That is, the
large scattering lengths arise from the 'cancelation' between $C^
{-1}_0$ and $J_0$, which must be effective measures of certain
objects in the pionfull theory. Hence, it is intriguing to extract
the pionless parameters like $J_0$ from pionfull theory as we did
for pionless couplings, i.e., to calculate $J_0$ from mapping
perspective. We will pursue such studies in future. The above
perspective should be generally true in the EFT descriptions of
many low-energy systems and useful for renormalization in various
EFT contexts.
\section{Prospective studies and summary} So far we have just
performed some primary analysis about the pionfull effective theory
in mapping perspective. Obviously, there are a lot more works to be
done in the future. (1) Generically, pionless couplings without
derivatives take the following form: $C_{0}+C_{0\tau}\bm{\tau}_1
\cdot\bm{\tau}_2+C_{0\sigma}\bm{\sigma}_1\cdot\bm{\sigma}_2+C_{0
\sigma\tau}\bm{\sigma}_1\cdot\bm{\sigma}_2\ \bm{\tau}_1\cdot\bm
{\tau}_2$ or $\tilde{C}_{0}+\tilde{C}_{0\sigma}\bm{\sigma}_1\cdot
\bm{\sigma}_2$ after using Fierz transformation. Firm conclusions
about these constants could only be drawn after higher loop diagrams
(with more iterations) are extensively studied. The same is true for
higher contact couplings. Moreover, it is also interesting to study
the extraction of the parameters like $J_0$ from pion-exchange
diagrams. From such studies, we could learn more about the pionfull
effective theory for nuclear forces and pin down the scenario for
pionless theory as a byproduct. (2) It remains to see if the
foregoing renormalization strategy could work out at higher orders
and/or in nonperturbative regime. It is also interesting to see if
the observation that power divergences are merely artefact of
non-relativistic truncation could be developed into efficient
working rules for calculations in non-relativistic regimes. In
particular, it would be interesting to see how this observation and
the mapping perspective could be applied to $3$N or multi-body
nuclear forces. (3) Recently, the IR enhancement in pion-exchange
diagrams has been exploited to study $N\bar{N}$ systems near
production threshold in Refs.\cite{JPMa1,JPMa2} using an effective
field theory similar to that for $NN$ system. It will be interesting
to explore the detailed mechanism in such effective theories and
related systems to gain further insights into pionfull effective
theories.

As far as OPE is concerned, the infrared enhancement is the main
driving force for working in nonperturbative regimes. It is also the
driving force for developing approaches that incorporate this
enhancement to various degrees. For example, infrared enhancement is
at least partially incorporated into the low-momentum effective
potentials constructed by integrating out modes above the scale
$\Lambda\approx2.1$ fm$^{-1}$\cite{low-k V}, as this benchmark scale
sits right in the middle of the pionfull region $\tilde{U}_{(\pi)}$:
$\frac{m_\pi+\Lambda_{(\pi)}}{2}\approx3m_\pi\approx2.1$ fm$^{-1}$,
very close to the enhancement scale $\sqrt{M_Nm_\pi}\approx 1.8$
fm$^{-1}$. Furthermore, if one wish to remove the infrared enhanced
items from iterated OPE, then: (1) The contact couplings in pionfull
theory must be promoted up to absorb such enhanced items; (2) The
iteration diagrams must be accordingly modified to avoid double
counting. Aside from what we discussed above, there might be other
sources of intricacy in pionfull effective theory, to name one, the
nature of sigma meson\cite{return-sigma1,return-sigma2} and its
couplings to pions and nucleons may also play some unknown roles in
the pionfull effective theory for nuclear forces.

In summary, we performed a primary analysis of the mapping between
pionfull and pionless effective field theories for nuclear forces.
The $2N$-reducible component of the planar box diagram was shown to
provide the dominant and yet definite contribution to the pionless
coupling $C_0$ in comparison with $2N$-irreducible components. This
anomalous mapping is due to the enhancement generated by the
low-lying nucleon poles and also happens in the BKV prescription of
OPE. As a byproduct, a simple strategy for renormalizing the
pionfull theory emerged from our interactive use of relativistic
and non-relativistic formulations. Prospective studies of related
issues are addressed.
\section*{Acknowledgement}The author wishes to express his deep
gratitude to X.-Y. Li (ITP, CAS), F. Wang (Nanjing U), Y. Jia (IHEP,
CAS), H.-Y. Jin (Zhejiang U), G.-H. Zhu (Zhejiang U) and Dr. Chen Wu
(SINAP, CAS) for their encouragements and helps. Helpful
communications with Dr. S. Nakamura (Osaka U) are also acknowledged.
We are grateful to the anonymous referee for his suggestions that
improve the presentation of our manuscript. This work is supported
in part by the Ministry of Education of China.
\appendix
\section{}
It is straightforward to see that the planar box diagram reduces to
the following integrals when external momenta are zero:\bea{T}^{(
\texttt{\tiny pb})}(\bm{0},\bm{0})=\frac{ig^4_A}{16f^4_\pi}(3-2\bm{
\tau}_1\cdot\bm{\tau}_2)\left[4M^2_NI_0-I_2+16M^4_NI_{2+-}-4M^2_N(I
_{2+}+I_{2-})\right],\eea where\bea\label{AppI2+-}{I}_n\equiv\int
\frac{d^4l}{(2\pi)^4}\frac{l_0^n}{A_\pi^2}\ (n=0,2),\ I_{2+-}\equiv
\int\frac{d^4l}{(2\pi)^4}\frac{l^2_0}{A^2_\pi A_+A_-},\ I_{2\pm}
\equiv\int\frac{d^4l}{(2\pi)^4}\frac{l^2_0}{A^2_\pi A_\pm},\eea with
${A}_\pi\equiv l^2-m^2_\pi+i\epsilon,\ A_\pm\equiv l^2\pm2M_Nl_0+i
\epsilon.$ Here, $I_{2+-}$ is definite, $I_0$ and $I_{2\pm}$ at most
carry logarithmic divergence. It is evident that $I_2$ only involve
pion propagators, so to stay chiral, the quadratic divergence
$\sim\frac{-i\Lambda^2}{2(4\pi)^2}$ in $I_2$ should be subtracted
together with the chiral divergences if one works in a covariant
cutoff scheme. Thus, for covariant form of chiral perturbation
theory, it is simpler to work with dimensional scheme:\bea{I}_0&=&
\frac{i}{(4\pi)^2}\left[\Gamma(\epsilon)-\ell_{\pi}\right],\quad{I}
_2=\frac{im^2_\pi}{2(4\pi)^2}\left[\Gamma(\epsilon)+1-\ell_{\pi}
\right],\\I_{2\pm}&=&\frac{i}{(8\pi)^2}\left[\Gamma(\epsilon)+1
-\ell_{N}+\frac{6\varrho+(3-4\varrho)\ln\varrho}{2\varrho^2}+\frac{
(3-10\varrho)\arctan\sqrt{4\varrho-1}}{\varrho^2\sqrt{4\varrho-1}}
\right],\\\label{I2+-}I_{2+-}&=&\frac{i}{(8\pi)^2}\frac{1}{M^2_N}
\left[\frac{(1-\varrho)\ln\varrho}{\varrho}+2+\frac{(2-6\varrho)
\arctan\sqrt{4\varrho-1}}{\varrho\sqrt{4\varrho-1}}\right],\eea
with\bea\ell_{\pi}\equiv\ln\frac{m^2_\pi}{\mu^2},\ \ell_N\equiv\ln
\frac{M_N^2}{\mu^2},\ \varrho\equiv\frac{M^2_N}{m^2_\pi}.\eea

\end{document}